%% file: errata.tex
\documentclass[twoside,11pt]{article}

\usepackage{amsmath}
\usepackage{amssymb}
\usepackage[boxruled,vlined,nokwfunc]{algorithm2e}
\usepackage{graphics,graphicx}
\usepackage{color}
\usepackage{multirow}
\usepackage{url}

\newcommand{\PIVOT}{\mbox{\sc Pivot}}
\newcommand{\RANDOM}{\mbox{\sc Random}}
\newcommand{\IMPROVE}{\mbox{\sc Improve}}

\usepackage{ifpdf}
\ifpdf
\else
\fi

\include{./macros_random_facet_shortest_paths}

\begin{document}

\title{\textbf{Errata for:} A subexponential lower bound for the \\Random Facet algorithm for Parity Games}

\author{Oliver Friedmann\thanks{Department of Computer Science,
University of Munich, Germany. E-mail: {\tt
  Oliver.Friedmann@gmail.com}.}\\
 \and
Thomas Dueholm Hansen\thanks{Department of Computer Science,
Aarhus University, Denmark. E-mail:
{\tt tdh@cs.au.dk}.}\\
\and
Uri Zwick\thanks{School of Computer Science, Tel Aviv University,
  Israel. E-mail: {\tt zwick@tau.ac.il}.} }
\date{}

\maketitle

\begin{abstract}\noindent%
In Friedmann, Hansen, and Zwick (2011) we claimed that the expected number of pivoting steps performed by the $\RandomFacet$ algorithm of Kalai and of Matou{\v{s}}ek, Sharir, and Welzl is equal to the expected number of pivoting steps performed by $\RandomFacet^*$, a variant of $\RandomFacet$ that bases its random decisions on one random permutation. We then obtained a lower bound on the expected number of pivoting steps performed by $\RandomFacet^*$ and claimed that the same lower bound holds also for $\RandomFacet$.
Unfortunately, the claim that the expected numbers of steps performed by $\RandomFacet$ and $\RandomFacet^*$ are the same is false. We provide here simple examples that show that the expected numbers of steps performed by the two algorithms are not the same.
\end{abstract}

\section{Introduction}

The $\RandomFacet$ algorithm was introduced by Kalai \cite{Kalai92,Kalai97} and by Matou{\v{s}}ek, Sharir, and Welzl \cite{MaShWe96}. It is a randomized pivoting rule for the simplex algorithm that solves linear programs, and more general \emph{LP-type problems}, in expected time $2^{O(\sqrt{(m-d)\log d})}$, where $m$ is the number of inequalities\footnote{It is customary to denote the number of inequalities by $n$, but we use $m$ to be consistent with the notation for graphs and shortest paths.} and $d$ is the dimension. Kalai's formulation of the algorithm works recursively as follows: Select a uniformly random facet containing the current vertex. Recursively find the optimal solution within the selected facet. If possible, perform an improving pivot from the resulting vertex and repeat. Otherwise return the vertex as the solution.

In Friedmann, Hansen, and Zwick \cite{FriedmannHansenZwick/SODA11} we proved a $2^{\tilde\Omega(\sqrt{m})}$ lower bound for the expected number of pivots performed by a modified variant, $\RandomFacet^*$, of the $\RandomFacet$ algorithm. The lower bound was proved for \emph{parity games}, which are of LP-type (see \cite{Halman07}). $\RandomFacet^*$ takes as input a fixed permutation of the facets and always selects the first available facet according to this permutation. We claimed that $\RandomFacet^*$ performs the same expected number of pivots as $\RandomFacet$ when the permutation is chosen uniformly at random. This would imply that our lower bound also holds for the original $\RandomFacet$ algorithm. The claim is incorrect, however, and in this errata we explain the error. The main result of \cite{FriedmannHansenZwick/SODA11} is thus flawed.

In Friedmann, Hansen, and Zwick \cite{FriedmannHansenZwick/STOC11} we transferred the lower bound construction from the setting of parity games to the setting of \emph{Markov decision processes}. This proves a lower bound for the $\RandomFacet^*$ algorithm for Markov decision processes, and consequently for linear programming. Again, due to our mistake, the lower bound does not hold for the original $\RandomFacet$ algorithm.

It should be stressed that the error is unrelated to the lower bound construction itself; it only involves the relationship between the $\RandomFacet$ algorithm and the modified version, $\RandomFacet^*$. In particular, the lower bound for $\RandomFacet^*$ remains correct. Also, the main focus of \cite{FriedmannHansenZwick/STOC11} was to prove a lower bound for the $\RandomEdge$ pivoting rule, and this work remains unaffected.

The error was pointed out briefly in \cite{Hansen12}, and an alternative proof of a $2^{\tilde\Omega(\sqrt[3]{m})}$ lower bound for the $\RandomFacet$ algorithm was given. Note that this lower bound is weaker than the $2^{\tilde\Omega(\sqrt{m})}$ lower bound we originally claimed in \cite{FriedmannHansenZwick/SODA11,FriedmannHansenZwick/STOC11}. The proof in \cite{Hansen12} was presented for Markov decision processes. In \cite{FriedmannHansenZwick/2014} we transfer the lower bound construction to the setting of shortest paths where we again prove a $2^{\tilde\Omega(\sqrt[3]{m})}$ lower bound for $\RandomFacet$, and a $2^{\tilde\Omega(\sqrt{m})}$ lower bound for $\RandomFacet^*$.

In this errata we explain our error in detail and give examples where the expected running times of $\RandomFacet$ and $\RandomFacet^*$ differ. We give a short introduction of the algorithms in Section~\ref{sec:algorithms}. To simplify the presentation we restrict our attention to shortest paths problems. In particular we adopt the notation from \cite{FriedmannHansenZwick/2014}. In Section \ref{sec:slower} we give an example where $\RandomFacet^*$ requires more pivots in expectation than $\RandomFacet$, and in Section \ref{sec:faster} we give an example where the opposite is the case. It is not difficult to obtain similar examples for parity games.

\section{The $\RandomFacet$ and $\RandomFacet^*$ algorithms}\label{sec:algorithms}

Let $G=(V,E,c)$ be a weighted directed graph, where $c:E\to\RR$ is a \emph{cost} function defined on its edges. Let $\TT\in V$ be a designated \emph{target} vertex. We let $n=|V\setminus\{\TT\}|$ and $m=|E|$ be the number of vertices, not counting the target, and edges in~$G$, respectively. We are interested in finding a tree of shortest paths from all vertices to~$\TT$.\footnote{To maintain consistency with \cite{FriedmannHansenZwick/SODA11,FriedmannHansenZwick/STOC11} it is more convenient for us to work with the \emph{single-target} version of the shortest paths problem, rather than the more standard \emph{single-source} version.}

Let $B \subseteq E$ be a tree containing directed paths from all vertices to~$\TT$. For some vertex $v \in V$, let $d_B(v)$ be the distance from $v$ to $\TT$ in the tree $B$. The $\RandomFacet$ algorithm takes two arguments: a set of edges $F \subseteq E$ and a tree $B\subseteq F$. It computes the tree of shortest paths for the subgraph defined by $F$ as follows. It picks a uniformly random edge $e \in F \setminus B$, removes $e$ from $F$, and finds the optimal tree $B'$ for the resulting subgraph defined by $F \setminus \{e\}$. It then checks whether including $e=(u,v)$ in $B'$ improves the solution, i.e., whether the path from $u$ to $\TT$ that starts with the edge $e$, which is currently not in the tree, is shorter than the path from $u$ to $\TT$ in the tree. If the path is shorter, then $e$ is exchanged with the edge $e'$ emanating from $u$ in $B'$, and a second recursive call is made with $F$ and $B'' = B' \cup \{e\} \setminus \{e'\}$. Note that updating the tree in this way corresponds to an improving pivot of the simplex algorithm. We assume for simiplicity that $G$ does not contain negative cycles. If the path is not shorter, then $B'$ is a tree of shortest paths for $F$.

Pseudo-code for \RandomFacet\ is given on the left of Figure~\ref{F-RandomFacet}. The first argument~$F \subseteq E$
is the set of available edges. Initially $F=E$. The second argument~$B$ is the current tree.
The call $\IMPROVE(B',e)$ checks whether $e$ can be used to improve $B'$.
The call $\PIVOT(B',e)$ returns the tree obtained from pivoting with $e$, i.e., it exchanges $e$ with some $e'\in B'$.

\begin{figure}[t]
\begin{center}
\parbox{3.2in}{
\SetAlgoFuncName{Algorithm}{anautorefname}
\begin{function}[H]
\DontPrintSemicolon
\SetAlgoRefName{}
\eIf{$F =B$}
{\Return{$B$}\;}
{
    $e\gets \RANDOM(F\setminus B)$ \;
    $B' \gets \RandomFacet(F \setminus\{e\},B)$ \;
    \eIf{$\IMPROVE(B',e)$}
    {
        $B'' \gets \PIVOT(B',e)$ \;
        \Return{$\RandomFacet(F,B'')$}\;
    }
    {
      \Return{$B'$}\;
    }
}
\caption{\RandomFacet($F,B$)}
\end{function}
}
\hspace*{10pt}
\parbox{3.2in}{
\SetAlgoFuncName{Algorithm}{anautorefname}
\begin{function}[H]
\DontPrintSemicolon
\SetAlgoRefName{}
\eIf{$F =B$}
{\Return{$B$}\;}
{
    $e\gets \argmin_{e' \in F \setminus B} \sigma(e')$ \;
    $B' \gets \RandomFacet^*(F \setminus\{e\},B,\sigma)$ \;
    \eIf{$\IMPROVE(B',e)$}
    {
        $B'' \gets \PIVOT(B',e)$ \;
        \Return{$\RandomFacet^*(F,B'',\sigma)$}\;
    }
    {
      \Return{$B'$}\;
    }
}
\caption{$\RandomFacet^*$($F,B,\sigma$)}
\end{function}
}
\end{center}
\caption{The $\RandomFacet$ and $\RandomFacet^*$ algorithms.}\label{F-RandomFacet}
\end{figure}

The $\RandomFacet^*$ algorithm is identical to the $\RandomFacet$ algorithm, except that it takes as an additional argument a permutation $\sigma: E \to \{1,\dots,|E|\}$ of the edges and always removes the first available edge according to this permutation. Pseudo-code for $\RandomFacet^*$ is shown on the right of Figure~\ref{F-RandomFacet}. Note that $\RandomFacet^*$ is a deterministic algorithm. We are however interested in the case when the permutation $\sigma$ is chosen uniformly at random.

Let $f(F,B)$ be the expected number of pivots performed by $\RandomFacet(F,B)$, and let $f^*(F,B)$ be the expected number of pivots performed by $\RandomFacet^*(F,B,\sigma)$ when $\sigma$ is uniformly random.
In Lemma 4.1 in \cite{FriedmannHansenZwick/SODA11} we claim that $f(F,B) = f^*(F,B)$ for all $F \subseteq E$ and all trees $B \subseteq F$. This claim is false, and we next show where the mistake was made.

Let $B$ and $F$ be given where $B \subseteq F \subseteq E$.
Suppose $\sigma$ is a permutation of $F$ such that the elements of $F\setminus B$ are ordered uniformly at random. Let $e = \argmin_{e' \in F \setminus B} \sigma(e')$ be the first available edge from $F \setminus B$ according to $\sigma$. At this stage, selecting $e$ according to $\sigma$ is equivalent to selecting $e \in F\setminus B$ uniformly at random. Let $B'$ be the optimal tree for $F\setminus \{e\}$, and assume that $e$ improves $B'$ such that $\RandomFacet$ and $\RandomFacet^*$ perform a second recursive call with $B'' = \PIVOT(B',e)$. Note that we may assume that $B'$ is uniquely determined by $F$ and $e$. In \cite{FriedmannHansenZwick/SODA11} we incorrectly assume that if $\sigma$ is uniformly random then the elements of $F\setminus B''$ are also ordered uniformly at random, and it is this assumption that is the source of our error.

In Sections \ref{sec:slower} and \ref{sec:faster} we give two concrete examples in which $f(F,B) < f^*(F,B)$ and $f(F,B) > f^*(F,B)$, respectively. The examples are for shortest paths. In the remainder of this section we consider an abstract example that illustrates why the permutation is not uniformly random for the second recursive call.

Suppose $F = \{1,2,3\}$ and $B = \{1\}$. For some permutation $\sigma$, we write $a \prec b$ as shorthand notation for $\sigma(a) < \sigma(b)$. There are six permutations of $F$:
\begin{align*}
&1 \prec 2 \prec 3 \quad\quad\quad
2 \prec 1 \prec 3 \quad\quad\quad
3 \prec 1 \prec 2\\
&1 \prec 3 \prec 2 \quad\quad\quad
2 \prec 3 \prec 1 \quad\quad\quad
3 \prec 2 \prec 1
\end{align*}
As there is an equal number of permutations in which $2\prec 3$ and $3\prec 2$, the first random choice made by $\RandomFacet^*$ is uniform. However, if, for example, $2\prec 3$, then only three equally likely permutations remain:
\begin{align*}
1 \prec 2 \prec 3\quad\quad\quad
2 \prec 1 \prec 3\quad\quad\quad
2 \prec 3 \prec 1
\end{align*}
Now, if $B'' = \{2\}$, then the distribution of the elements from $F\setminus B'' = \{1,3\}$ is no longer uniform for the second recursive call of $\RandomFacet^*$. We now have $1 \prec 3$ with probability $2/3$, and
$3 \prec 1$ with probability $1/3$. Thus, elements that leave $B$ are more likely to be picked sooner in the second recursive call.

\section{Example: $\RandomFacet^*$ can be slower than $\RandomFacet$}\label{sec:slower}

\begin{figure}[t]
\begin{center}
\includegraphics[scale=1]{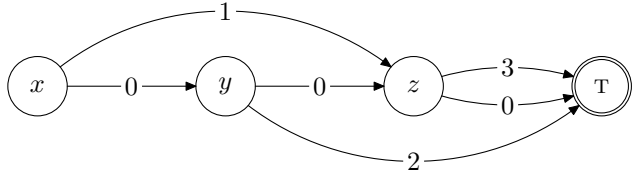}
\end{center}
\caption{A shortest paths problem in which the expected number of improving switches performed by \RandomFacet\ and $\RandomFacet^*$ differ.}\label{fig:example}
\end{figure}

Consider the simple example of a shortest paths problem given in Figure \ref{fig:example}. There are three non-terminal vertices $x,y,z$, each with two out-going edges. We refer to the edges leaving $x$ as $x_0$ and $x_1$, where the cost of $x_0$ is 0 and the cost of $x_1$ is positive. We refer to the other edges in a similar fashion such that the set of edges is $E = \{x_0,x_1,y_0,y_1,z_0,z_1\}$. Note that a tree $B = \{x_i,y_j,z_k\} \subseteq E$ can be viewed as a binary string $ijk$ of length 3. In particular the set of trees can be viewed as vertices of a cube, and in fact the corresponding linear program is combinatorially equivalent to a cube.

Consider two neighboring vertices of the cube. Moving from one vertex to the other corresponds to performing a pivot step, and we orient the edge in the improving direction. When all edges are oriented, the resulting orientation is known as an \emph{acyclic unique sink orientation} (see, e.g., \cite{SzWe01}). Note that a set of edges $F$ defines a corresponding \emph{face} that contains the trees in $F$. In every such face there is a unique optimal solution, which means that in the subcube there is a unique vertex where all the edges are incoming. Such a vertex is called a \emph{sink}.

\begin{figure}[t]
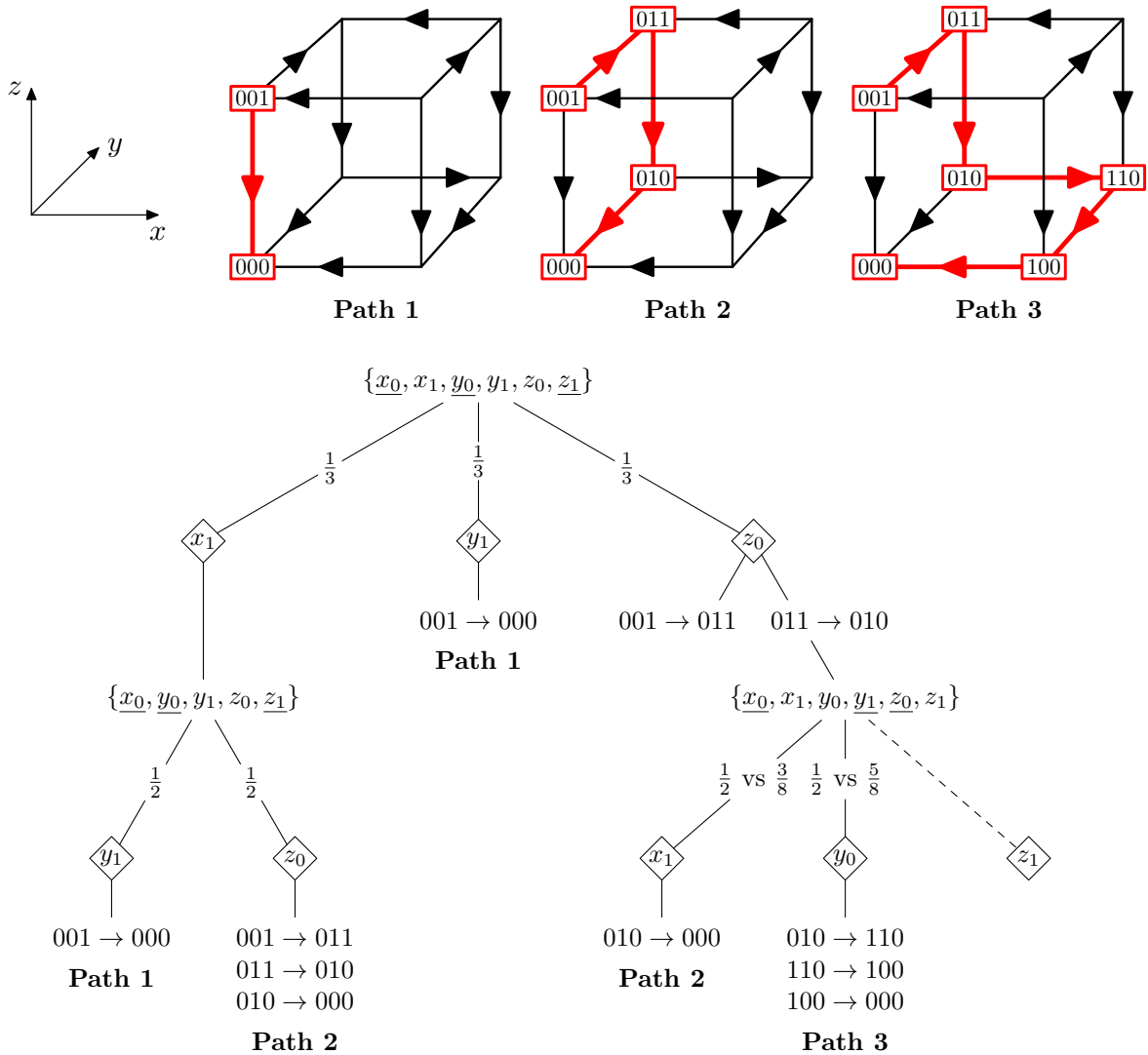

\begin{center}
\parbox{1.2in}{
\center
\includegraphics[scale=2.3]{errata.2}
}
\hspace*{5pt}
\parbox{1.5in}{
\center
\includegraphics[scale=2.3]{errata.3}
}
\hspace*{5pt}
\parbox{1.5in}{
\center
\includegraphics[scale=2.3]{errata.4}
}
\hspace*{5pt}
\parbox{1.5in}{
\center
\includegraphics[scale=2.3]{errata.5}
}

\vspace*{5pt}

\parbox{6in}{
\center
\includegraphics{errata.9}
\caption{Paths generated by the $\RandomFacet$ and $\RandomFacet^*$ algorithms when starting from 001.}\label{fig:slower}
}
\end{center}
\end{figure}

The orientation corresponding to the graph in Figure \ref{fig:example} is shown in Figure \ref{fig:slower}. For the example we start with the tree $B=\{x_0,y_0,z_1\}$, which corresponds to the vertex 001. The optimal solution is the vertex 000, and there are three paths (sequences of pivots) leading from 001 to 000. These paths are shown at the top of Figure \ref{fig:slower}. The vertex 000 is in the lower left corner of the cube, and the cube is oriented according to the axes shown on the left.

The bottom of Figure \ref{fig:slower} shows the computations performed by \RandomFacet\ and $\RandomFacet^*$. The set $F = \{x_0,x_1,y_0,y_1,z_0,z_1\}$ is the set of edges given as an argument to the calls $\RandomFacet(F,B)$ and $\RandomFacet^*(F,B,\sigma)$. The elements in $B$ are underlined, and $\sigma$ is assumed to be uniformly random. The labelled squares indicate the edges that are chosen by the algorithms. The probabilities of picking the edges are shown as well. In most cases the probabilities are the same, however, the label $\frac{1}{2} ~\text{vs}~ \frac{3}{8}$ means that $\RandomFacet$ picks the edge with probability $\frac{1}{2}$ and $\RandomFacet^*$ picks the edge with probability $\frac{3}{8}$. The lines below the squares indicate recursive calls with the chosen edge. The first recursive call is shown on the left, and if a second recursive call is made, it is shown on the right. Lines corresponding to second recursive calls are labelled by the pivot that is made. Recall that the vertex returned by the first recursive call is uniquely determined. When only a single sequence of pivots can be generated in a subtree this sequence is shown instead of the subtree itself. It is also shown which of the paths the sequence corresponds to. For a selection of edges, the sequence of pivots for the generated path can be read in a pre-order traversal of the tree.

The line to $z_1$ is dashed because removing this edge does not affect the behavior of the algorithms. More precisely there is no path from 010 that leads back to the $z_1$ facet. (In fact, the edge that leaves a tree during a pivot can in general never appear in the corresponding second recursive call, and it may be viewed as if it has been removed.) We instead combine the probabilities that should have appeared after $z_1$ with those at the same recursive call as $z_1$.

Consider the path in the computation tree that first picks $z_0$ and then picks $y_0$.
In order for the call
\[
\RandomFacet^*(\{x_0,x_1,y_0,y_1,z_0,z_1\},\{x_0,y_0,z_1\},\sigma)
\]
to realize this path, the permutation $\sigma$ must satisfy $z_0 \prec x_1$, $z_0 \prec y_1$, and $y_0 \prec x_1$. There are 150 permutations that satisfy these restrictions, which means that the path is generated with probability $150/6! = 5/24$, and the edge is picked with probability $5/8$, given that the call is made. The permutations can be obtained by adding $x_0$ and $z_1$ to the following five permutations:
\begin{align*}
&y_0 \prec z_0 \prec x_1 \prec y_1 \quad\quad\quad
z_0 \prec y_0 \prec x_1 \prec y_1 \quad\quad\quad
z_0 \prec y_1 \prec y_0 \prec x_1 \\
&y_0 \prec z_0 \prec y_1 \prec x_1 \quad\quad\quad
z_0 \prec y_0 \prec y_1 \prec x_1
\end{align*}
The same choice is made with probability $1/2$ by $\RandomFacet$. Note that $y_0$ was part of the original tree $B$, and as illustrated in Section \ref{sec:algorithms} this means that it is more likely to be picked during a second recursive call, which is exactly the behavior we observe here. The resulting path, \textbf{Path 3}, is longer than the path, \textbf{Path 2}, generated when $x_1$ is picked. Thus the expected running time is worse for $\RandomFacet^*$ than for $\RandomFacet$. To be exact, the expected number of pivots made by $\RandomFacet$ is
\[
f(E,\{x_0,y_0,z_1\}) ~=~ \frac{1}{3}\left(\frac12 + \frac32\right) + \frac13 +
\frac{1}{3}\left(\frac32 + \frac52\right) ~=~ \frac73~\approx~ 2.333~,
\]
and
the expected number of pivots made by $\RandomFacet^*$ is
\[
f^*(E,\{x_0,y_0,z_1\}) ~=~ \frac{1}{3}\left(\frac12 + \frac32\right) + \frac13 +
\frac{1}{3}\left(\frac98 + \frac{25}{8}\right) ~=~ \frac{29}{12} ~\approx~ 2.417~.
\]

\section{Example: $\RandomFacet^*$ can be faster than $\RandomFacet$}\label{sec:faster}

\begin{figure}[t]
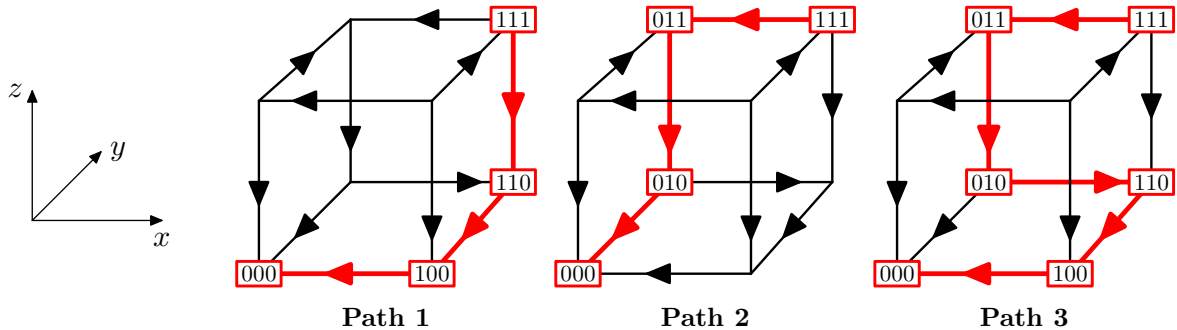
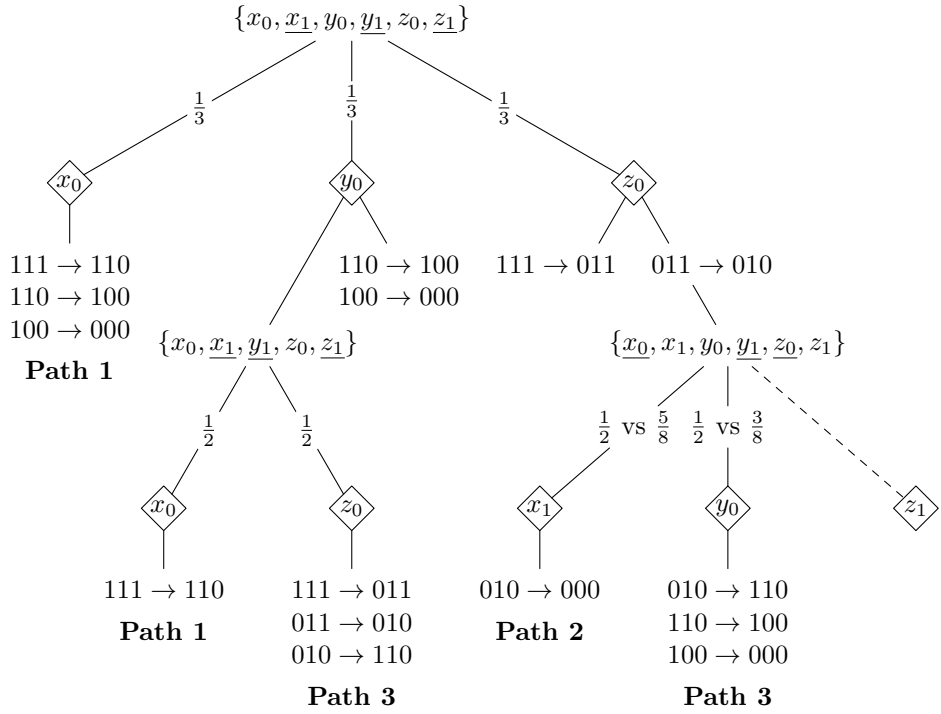

\begin{center}
\parbox{1.2in}{
\center
\includegraphics[scale=2.3]{errata.2}
}
\hspace*{5pt}
\parbox{1.5in}{
\center
\includegraphics[scale=2.3]{errata.6}
}
\hspace*{5pt}
\parbox{1.5in}{
\center
\includegraphics[scale=2.3]{errata.7}
}
\hspace*{5pt}
\parbox{1.5in}{
\center
\includegraphics[scale=2.3]{errata.8}
}

\vspace*{5pt}

\parbox{6in}{
\center
\includegraphics{errata.10}
\caption{Paths generated by the $\RandomFacet$ and $\RandomFacet^*$ algorithms when starting from 111.}\label{fig:faster}
}
\end{center}
\end{figure}

We next give an example for which $\RandomFacet^*$ is faster than $\RandomFacet$. In fact the example is again for the graph shown in Figure~\ref{fig:example}, and we therefore use the same notation as in Section~\ref{sec:slower}. The only difference is that we now start from the vertex 111. The resulting paths and computation tree are shown in Figure \ref{fig:faster}. There are again only three paths from 111 to 000.

Consider the path in the computation tree that first picks $z_0$ and then picks $x_1$. Note that $x_1$ is part of the original tree $B= \{x_1,y_1,z_1\}$, and we would therefore expect this choice to be made with higher probability for $\RandomFacet^*$ than for $\RandomFacet$, which is exactly the case.
In order for the call
\[
\RandomFacet^*(\{x_0,x_1,y_0,y_1,z_0,z_1\},\{x_1,y_1,z_1\},\sigma)
\]
to realize the path, the permutation $\sigma$ must satisfy $z_0 \prec x_0$, $z_0 \prec y_0$, and $x_1 \prec y_0$. There are 150 permutations that satisfy these restrictions, which means that the path is generated with probability $150/6! = 5/24$, and the edge is picked with probability $5/8$, given that the call is made. The permutations can be obtained by adding $y_1$ and $z_1$ to the following five permutations:
\begin{align*}
&x_1 \prec z_0 \prec x_0 \prec y_0 \quad\quad\quad
z_0 \prec x_1 \prec x_0 \prec y_0 \quad\quad\quad
z_0 \prec x_0 \prec x_1 \prec y_0 \\
&x_1 \prec z_0 \prec y_0 \prec x_0 \quad\quad\quad
z_0 \prec x_1 \prec y_0 \prec x_0
\end{align*}
The same choice is again made with probability $1/2$ by $\RandomFacet$. This time, the resulting path, \textbf{Path 2}, is shorter than the path, \textbf{Path 3}, generated when $y_0$ is picked. Thus the expected running time is better for $\RandomFacet^*$ than for $\RandomFacet$. To be exact, the expected number of pivots made by $\RandomFacet$ is
\[
f(E,\{x_0,y_0,z_1\}) ~=~ 1 + \frac{1}{3}\left(\frac32 + \frac52\right) +
\frac{1}{3}\left(\frac32 + \frac52\right) ~=~ \frac{11}{3}~\approx~ 3.667~,
\]
and
the expected number of pivots made by $\RandomFacet^*$ is
\[
f^*(E,\{x_0,y_0,z_1\}) ~=~ 1 + \frac{1}{3}\left(\frac32 + \frac52\right) +
\frac{1}{3}\left(\frac{15}{8} + \frac{15}{8}\right) ~=~ \frac{43}{12}~\approx~ 3.583~.
\]

\section*{Acknowledgement}

We would like to thank Bernd G{\"a}rtner, G\"{u}nter Rote and Tibor Szab\'{o} for various discussions on the \RandomFacet\ algorithm and its variants that helped us realize that the expected number of pivoting steps performed by \RandomFacet\ and $\RandomFacet^*$ are \emph{not} the same.

\bibliographystyle{abbrv}
\bibliography{../../random_facet_shortest_paths}

\end{document}

%% file: macros_random_facet_shortest_paths.tex
\newcommand{\RR}{{\mathbb R}}

\DeclareMathOperator*{\argmin}{argmin}

\newcommand{\mathsc}[1]{\mbox{\textsc #1}}

\newcommand{\TT}{{\mathsc t}}

\newcommand{\RandomFacet}{\mbox{\sc Random-}\allowbreak\mbox{\sc Facet}}

\newcommand{\RandomEdge}{\mbox{\sc Random-}\allowbreak\mbox{\sc Edge}}

\newcommand{\ignore}[1]{}

%% file: errata.bbl
\begin{thebibliography}{1}

\bibitem{FriedmannHansenZwick/SODA11}
O.~Friedmann, T.~D. Hansen, and U.~Zwick.
\newblock A subexponential lower bound for the random facet algorithm for
  parity games.
\newblock In {\em Proc.\ of 22nd SODA}, pages 202--216, 2011.

\bibitem{FriedmannHansenZwick/STOC11}
O.~Friedmann, T.~D. Hansen, and U.~Zwick.
\newblock Subexponential lower bounds for randomized pivoting rules for the
  simplex algorithm.
\newblock In {\em Proc.\ of 43th STOC}, pages 283--292, 2011.

\bibitem{FriedmannHansenZwick/2014}
O.~Friedmann, T.~D. Hansen, and U.~Zwick.
\newblock {R}andom-{F}acet and {R}andom-{B}land require subexponential time
  even for shortest paths.
\newblock {\em CoRR}, abs/1410.7530, 2014.

\bibitem{Halman07}
N.~Halman.
\newblock Simple stochastic games, parity games, mean payoff games and
  discounted payoff games are all {LP}-type problems.
\newblock {\em Algorithmica}, 49(1):37--50, 2007.

\bibitem{Hansen12}
T.~D. Hansen.
\newblock {\em Worst-case Analysis of Strategy Iteration and the Simplex
  Method}.
\newblock PhD thesis, Aarhus University, 2012.
\newblock Available at: \url{www.cs.au.dk/~tdh/papers/dissertation.pdf}.

\bibitem{Kalai92}
G.~Kalai.
\newblock A subexponential randomized simplex algorithm (extended abstract).
\newblock In {\em Proc.\ of 24th STOC}, pages 475--482, 1992.

\bibitem{Kalai97}
G.~Kalai.
\newblock Linear programming, the simplex algorithm and simple polytopes.
\newblock {\em Mathematical Programming}, 79:217--233, 1997.

\bibitem{MaShWe96}
J.~Matou{\v{s}}ek, M.~Sharir, and E.~Welzl.
\newblock A subexponential bound for linear programming.
\newblock {\em Algorithmica}, 16(4-5):498--516, 1996.

\bibitem{SzWe01}
T.~Szab{\'o} and E.~Welzl.
\newblock Unique sink orientations of cubes.
\newblock In {\em Proc.\ of 42th FOCS}, pages 547--555, 2001.

\end{thebibliography}
